# Material-Tailored Multispectral Magnetization via the Inverse Faraday Effect in Tunable Coaxial Nanoantenna

Roméo Zapata[1], Bleuenn Belz[1], Xingyu Yang[1,2], Ye Mou[3], Wajdi Chabani[1], Maria Sanz-Paz[1] and Mathieu Mivelle[1*]

[1]Sorbonne Université, CNRS, Institut des NanoSciences de Paris, INSP, F-75005 Paris, France

[2]School of Optoelectronic Engineering and Instrumentation Science, Dalian University of Technology, Dalian 116024, China.

[3]School of Electronic and Information Engineering, Ningbo University of Technology, No. 201, Fenghua Road, Jiangbei District, Ningbo, Zhejiang, China

*Corresponding authors:

mathieu.mivelle@sorbonne-universite.fr

**Abstract**

All-optical control of magnetization via the Inverse Faraday Effect presents a transformative pathway for ultrafast high-density data storage and nanoscale spintronics. However, unlocking its full potential requires moving beyond traditional gold-centric, narrowband constraints. In this study, we present a comprehensive theoretical framework evaluating IFE-driven static magnetic field generation across a diverse library of plasmonic materials, including gold, silver, copper, aluminum, and titanium nitride. Using a dimensionally tunable coaxial nanoantenna, we map the optomagnetic response across the deep-ultraviolet to near-infrared spectrum. Our findings reveal highly distinct, material-dependent behaviors: aluminum uniquely sustains robust magnetization in the deep-UV, whereas silver, copper, and gold exhibit superior efficiency at longer wavelengths. Crucially, we demonstrate that shifting the excitation to the near-infrared regime amplifies the generated magnetic field by several orders of magnitude, establishing these frequencies as critical for helicity-dependent all-optical switching. Furthermore, by resolving the local optical spin density, we elucidate how extreme subwavelength confinement drives synchronized drift currents, enabling efficient multispectral magnetization through localized surface plasmon resonances and higher-order Fabry-Perot modes. Strikingly, comparing these distinct modes reveals that structural geometry tunes the resonant wavelength, while the maximum achievable optomagnetic amplitude is fundamentally governed by the metal's intrinsic properties.

## Introduction

Modern information technology is built upon two foundational pillars: the ultrafast data transfer of optical communications and the robust, non-volatile nature of magnetic data storage. Over the past few decades, the drive to seamlessly integrate these domains has sparked immense interest in the all-optical control of magnetic order. Initiated by pioneering observations of ultrafast demagnetization using pulsed lasers and subsequent demonstrations of helicity-dependent all-optical switching, this rapidly growing field aims to manipulate magnetization at increasingly faster temporal scales and smaller spatial dimensions. Bridging photonics and magnetism thus offers a compelling pathway towards next-generation magneto-optic devices.

A promising non-thermal mechanism to achieve this all-optical control lies in the Inverse Faraday Effect (IFE). First predicted and observed in the 1960s[1,2] as a phenomenological counterpart to the conventional Faraday effect, the IFE is a fundamental light–matter interaction wherein circularly polarized light magnetizes a medium along its propagation axis, with the magnetization sign dictated by the light's helicity.[3–5] In non-magnetic metals, this near-instantaneous, non-linear process imparts continuous circular motion to conduction electrons [6–8]. By driving these non-dissipative circulating drift photocurrents, the IFE generates localized, stationary magnetic fields on subpicosecond timescales.[9–11] Grounded in early classical descriptions of plasma physics, [12,13] the IFE has proven to be a highly general light-matter interaction, offering fertile ground for new scientific breakthroughs. [14]

Crucially, recent advances in nanophotonics and plasmonics have fundamentally redefined the scope of the IFE by enabling the precise spatial engineering of optical spin density and local field gradients at the nanoscale.[15–17] By harnessing Localized Surface Plasmon Resonances (LSPR), plasmonic nanostructures can tightly confine and amplify optical fields.[18,19] This extreme light concentration drastically enhances the non-linear optical forces acting on the metal's conduction electrons.[20–22] When these amplified drift photocurrents are guided through rationally designed inductive geometries, matter can be magnetized at spatial scales fundamentally inaccessible to conventional electromagnets.[23–26]

However, while the theoretical foundations and experimental evidence for IFE generation in plasmonic nanostructures are well established, a critical bottleneck remains. Previous investigations into IFE-driven magnetization have predominantly focused on gold nanostructures operating within limited spectral windows.[27] Unlocking the full potential of all-optical magnetism requires a systematic numerical framework to deeply understand and optimize this opto-magnetic conversion across diverse material compositions and a much broader spectral range.

To address this gap, we present a comprehensive theoretical study using plasmonic coaxial nanoantennas as a highly versatile platform. While the inherent geometry of coaxial systems is well known for maximizing capacitive coupling and creating intense optical "hotspots"[19,28,29] its paramount value for our investigation lies in its capacity to support tunable Fabry-Perot resonances.[30-32] By adjusting the structure's thickness, these higher-order resonances enable us to continuously shift the optical response across the entire deep ultraviolet to near-infrared spectrum, providing a perfect testbed for evaluating different materials at their intrinsic optimal wavelengths.

Exploiting this broad spectral tunability, we deploy an advanced high-resolution numerical methodology to compute the non-linear IFE response across an extensive parameter space. This approach allows us to look beyond gold and systematically benchmark a broader library of materials, including Copper, Aluminum, Silver, and Titanium Nitride, against critical geometric variables (thickness $T$, inner diameter $D$, and excitation wavelength $\lambda$). Building on these large-scale computations, we construct a quantitative reference map that directly correlates material selection, geometry, and spectral window to optimize static magnetic field generation. Crucially, we highlight the distinct frequency-dependent responses of different metals and demonstrate how geometry (particularly thickness) acts synergistically with material permittivity to enable multispectral excitations. Ultimately, our findings establish a robust theoretical foundation for the rational design of highly efficient, material-tailored all-optical magnetic nanostructures.

## Results

To benchmark the multispectral performance of different plasmonic materials, we systematically evaluated the out-of-plane IFE magnetization efficiency across a carefully controlled geometric and spectral parameter space using the coaxial nanoantenna platform (**Figure 1**). To isolate material-specific behaviors while maintaining practical nanofabrication feasibility via standard electron beam lithography, the annular gap width was fixed at $G = 30$ nm (**Figure 1a**). Within this framework, we performed an exhaustive geometric scan encompassing a diverse library of material compositions, specifically Gold, Copper, Aluminum, Silver, and Titanium Nitride (TiN). For each material, we systematically varied the inner core diameter ($D$) from $20$ nm to $300$ nm in $25$ nm increments across multiple structural thickness tiers ($T = 50$ nm, $75$ nm, and $100$ nm, with extended evaluations up to $200$ nm, as defined in **Figure 1b**). This comprehensive parameter sweep spans the entire deep-UV to near-infrared spectrum ($200$–$2000$ nm), providing a rigorous reference map to evaluate how material permittivity and geometric tuning synergistically govern IFE-driven magnetic responses across ultraviolet, visible, and infrared regimes.

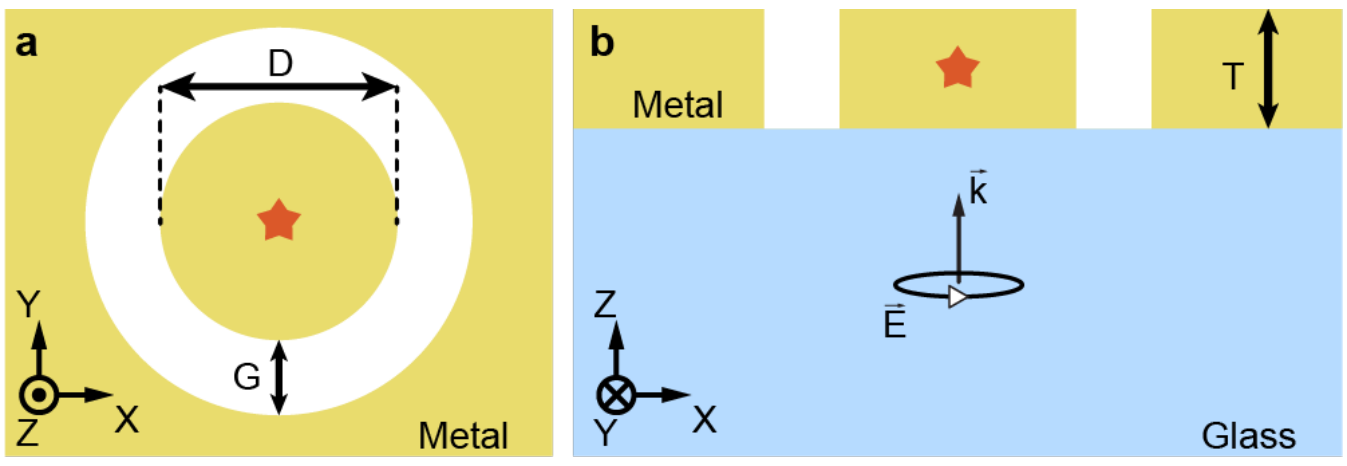

**Figure 1: Nanoscale coaxial structure for localized out-of-plane magnetic field generation.** (a) Top-view schematic in the XY plane defining the critical lateral geometrical parameters: the inner core diameter ($D$) and the annular gap width ($G$). (b) Cross-sectional schematic in the XZ plane defining the nanostructure thickness ($T$). The precise spatial location where the out-of-plane IFE magnetic field ($B_z(\lambda)$) is evaluated is indicated by a red star in both projections. The nanoantenna is excited by a broadband (200–2000 nm) effective right-handed circularly polarized (RCP) light source propagating along the $z$-axis. Details regarding the numerical implementation of this effective source are provided in the Supplementary Information.

Shifting focus to material composition provides critical insight into both the spectral versatility and the practical viability of the nanostructures. As logarithmically mapped in **Figure 2a–o**, the generated out-of-plane IFE magnetic field, $B_z(\lambda)$, reveals highly distinct behaviors dictated by the interplay between material permittivity and nanoscale geometry. Most notably, the inner core diameter ($D$) serves as a highly effective tuning mechanism, continuously shifting the fundamental LSPR mode across the optical spectrum.

Crucially, this spectral mapping validates a fundamental operational principle for future optomagnetic devices: the magnitude of the generated magnetic field is profoundly enhanced at longer wavelengths. Moving from the visible spectrum into the near-infrared (NIR) regime yields a drastic amplification of the IFE response, with $B_z$ intensities increasing by several orders of magnitude. This tremendous gain in efficiency makes the NIR spectrum inherently superior for practical applications, such as coupling these plasmonic nanoantennas to ferromagnetic or ferrimagnetic thin films for all-optical data storage. Operating at these longer wavelengths is therefore not merely a design option, but a critical prerequisite for achieving the high magnetic field strengths necessary for robust spintronic functionalities.

In this highly favorable NIR regime, the intrinsic optical properties of conventional metals, with the notable exception of Titanium Nitride, support intense resonant responses. Among them, Gold and Silver stand out as the absolute best performers, yielding the highest IFE efficiencies. Silver, in particular, demonstrates remarkable broadband versatility, maintaining strong coupling from the lower visible spectrum deep into the NIR (**Figure 2j–l**). However, Silver's rapid oxidation in ambient air presents a major experimental bottleneck, complicating its integration into reliable, long-term operational devices. In this context, Copper emerges as a highly strategic alternative. While its plasmonic performance at visible optical frequencies is underwhelming, it demonstrates a robust resonant response in the NIR (**Figure 2d–f**), solidifying its role as a cost-efficient surrogate for Gold in these longer wavelengths where the magnetic yield is maximal.

At the shorter end of the spectrum, Aluminum (**Figure 2g–i**) distinguishes itself through its unique capacity to sustain robust IFE magnetization deep into the ultraviolet region (down to $200$–$250$ nm). Coupled with its high natural abundance, Aluminum represents a highly attractive candidate for UV optomagnetism. Nevertheless, it exhibits a distinct, intrinsic performance drop around $800$ nm, a fundamental weakness directly tied to its specific optical properties and inter-band transitions in this spectral window. Meanwhile, Titanium Nitride (TiN), despite offering exceptional thermal stability and mechanical robustness, fails to support the intense NIR resonances seen in the other materials, yielding magnetic fields approximately one order of magnitude weaker across the entire spectrum (**Figure 2m–o**).

Furthermore, this extensive parameter sweep confirms that for thinner structures, the dominant magnetic-field peak arises almost exclusively from the fundamental LSPR. However, as the nanostructure thickness ($T$) reaches $100$ nm, common plasmonic materials begin to support a secondary set of higher-order resonance modes. Although less intense than the primary LSPR, these modes become clearly discernable in the UV–visible spectra (**Figure 2c, f, i, l**), highlighting the distinctive role of thickness in breaking simple scaling laws and enabling multispectral operation.

Building upon these spectral maps, we investigated how the inner core diameter ($D$) and material composition act synergistically as precise tuning parameters to shift the fundamental LSPR enhancement across the optical spectrum. To quantitatively illustrate this behavior, the scatter plots in **Figure 2p–r** compare the optimal geometric configurations for each material composition.

These panels, color-coded by the maximum out-of-plane magnetization intensity ($B_z$), map the direct relationship between the core diameter ($D$) and the corresponding resonant excitation wavelength for each nanostructure thickness tier ($T = 50$ nm, $75$ nm, and $100$ nm). Strikingly, Gold, Silver, and Copper exhibit a remarkably similar, nearly linear dependence between the resonant wavelength and the core size. This overlapping scaling behavior suggests a high degree of geometric interchangeability in device design; a nanoantenna geometry optimized for Gold could theoretically be swapped for Copper for near-infrared applications with minimal structural recalibration.

In stark contrast, Aluminum exhibits distinctly blue-shifted resonant wavelengths across all core diameters. This divergence from the noble metals fundamentally underpins its outstanding performance for IFE magnetization in the UV and blue spectral regions, where other metals typically suffer from severe optical damping. Ultimately, this visualization serves as a comprehensive quantitative design map. By directly correlating the required excitation wavelength with the necessary nanostructure geometry and material choice, it provides indispensable guidelines for optimizing precise nanofabrication protocols.

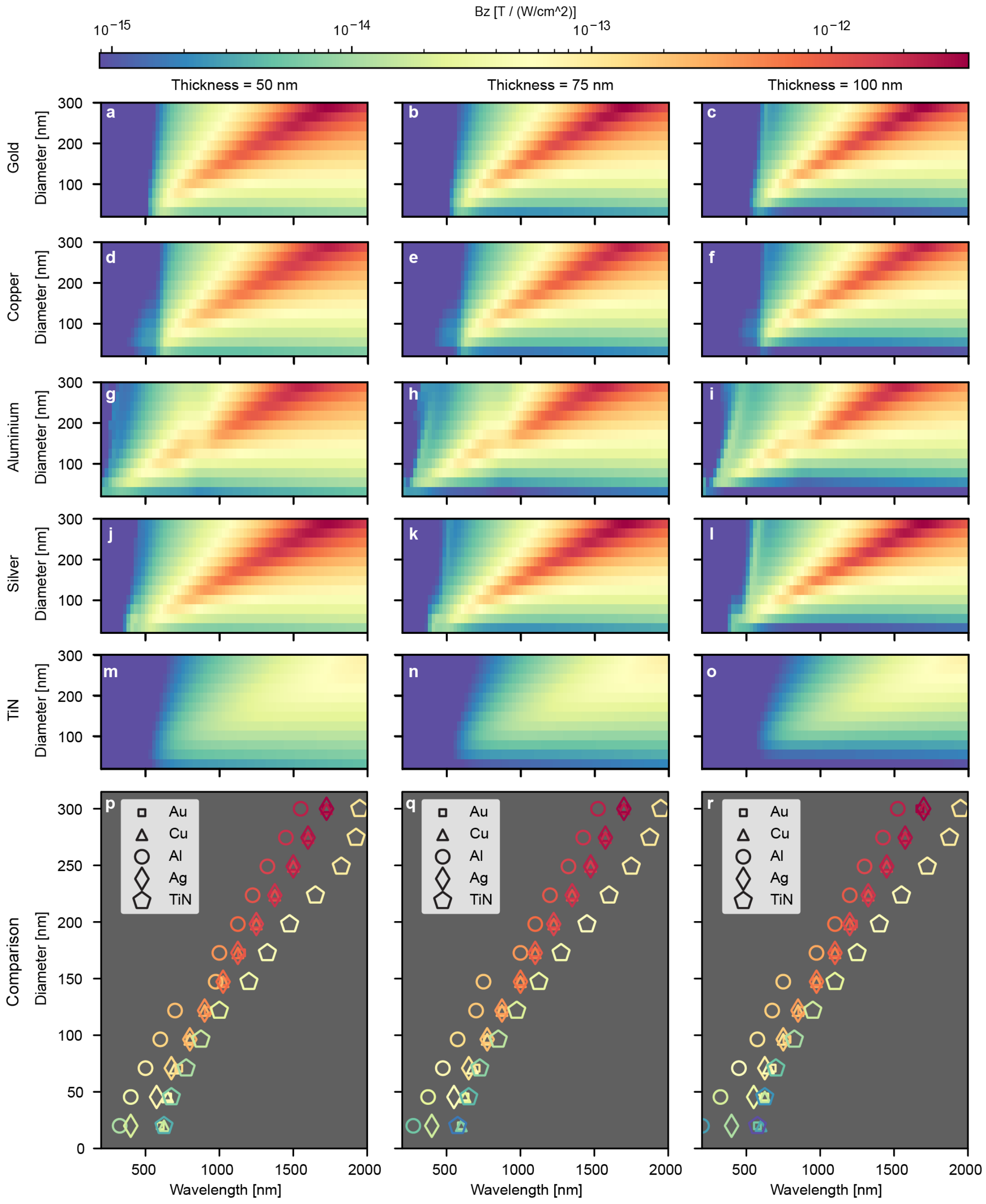

**Figure 2: Spectral selectivity of the IFE magnetic field across a diverse material library. (a–o)** Logarithmic maps of the induced out-of-plane IFE magnetic field spectra ($B_z(\lambda)$) demonstrating the wavelength and geometric dependence for five plasmonic materials: Gold (a–c), Copper (d–f), Aluminum (g–i), Silver (j–l), and Titanium Nitride (m–o). The parameter space spans the UV-to-NIR spectrum ($200 - 2000$ nm) as a function of the inner core diameter $D$ ($20 - 300$ nm) with a fixed annular gap $G = 30$ nm. For each material, the spectra are categorized by nanostructure thickness: $T = 50$ nm (left column: a, d, g, j, m), $75$ nm (middle column: b, e, h, k, n), and $100$ nm (right column: c, f, i, l, o). **(p–r)** Quantitative design maps extracting the maximum achievable IFE magnetic field ($B_z$) and the corresponding optimal excitation wavelength for each inner core diameter ($D$). These reference panels are categorized by thickness: (p) $50$ nm, (q) $75$ nm, and (r) $100$ nm

Moving beyond the fundamental LSPR, we explored the secondary resonant behaviors by expanding our analysis across a broader range of nanostructure thicknesses ($T = 50$ nm, $75$ nm, $100$ nm, and $200$ nm). For this detailed investigation, we selected a Gold coaxial nanoantenna with a large inner core diameter ($D = 300$ nm) as a representative model system for plasmonic metals (**Figure 3**). The spectral data reveal that increasing the nanostructure thickness not only amplifies the intensity of the generated out-of-plane magnetic field, but also induces a slight blue shift in the primary resonant wavelength.

Moreover, thicker structures support a greater number of resonant modes while simultaneously reducing the full-width at half-maximum (FWHM) of the resonance peaks. This spectral narrowing indicates an enhancement of the resonance quality factors ($Q$). By tracking these multiple modes, indexed from 1 to 6, we can observe how geometry dictates two distinct plasmonic regimes. First, all evaluated structures exhibit an intense IFE magnetic field distribution peaking at the primary LSPR wavelength near $1700$ nm (peaks 3–6), achieving high field intensities on the order of $10^{-11}$ $\text{T}/(\text{W/cm}^2)$. Second, structures reaching or exceeding $100$ nm in thickness trigger the emergence of tunable first-order Fabry-Perot (FP) modes spanning the visible-to-near-infrared spectrum ($500$–$1000$ nm, peaks 1 and 2, corresponding to thicknesses of $100$ nm and $200$ nm, respectively). Ultimately, the coexistence and geometric tunability of both LSPR and FP resonances confirm that manipulating nanostructure thickness is a powerful, highly predictable strategy for engineering magnetic fields across multiple spectral windows simultaneously.

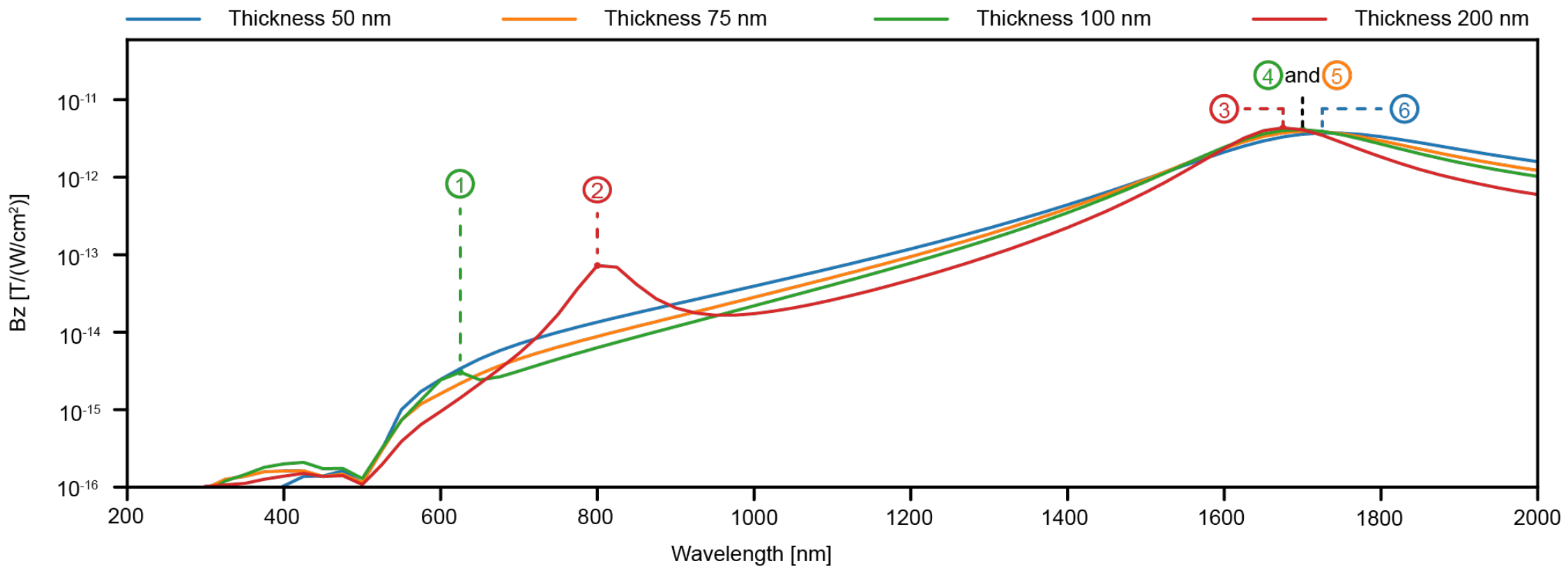

**Figure 3: Plasmonic Fabry Perot Resonances.** Spectra of the out-of-plane magnetic field component ($B_z$) generated by the IFE at the center of a Gold coaxial nanoantenna. The inner core diameter is fixed at $D = 300$ nm, while the nanostructure thickness is varied: $T =$

$50$ nm (blue), $75$ nm (orange), $100$ nm (green), and $200$ nm (red). Resonant peaks are indexed from 1 to 6 to highlight two distinct plasmonic regimes. Peaks 3–6 correspond to the primary Localized Surface Plasmon Resonance (LSPR) in the near-infrared region (near $1700$ nm), present across all evaluated thicknesses. Conversely, peaks 1 and 2 (between $500$ nm and $1000$ nm) correspond to first-order Fabry-Perot (FP) modes, which exclusively emerge in thicker structures ($T \geq 100$ nm).

To definitively validate the physical nature of these resonances, **Figure 4** maps the tightly confined near-field interactions. For the primary LSPR modes in the near-infrared, the cross-sectional analysis reveals that the optical energy is distributed homogeneously across the entire thickness of the coaxial aperture. Crucially, the resulting IFE magnetization is fundamentally driven by the local optical spin density ($S$), a vector quantity proportional to Im$(\mathrm{E}^* \times \mathrm{E})$ that characterizes the local polarization state and intrinsic spin angular momentum of the electromagnetic field at the nanoscale. Its out-of-plane component, $S_z$, directly maps the local ellipticity of light: positive values correspond to right-handed elliptical polarization, while negative values indicate left-handed elliptical polarization.

By mapping $S_z$ across the nanostructure, we observe a remarkable polarization gradient driven by extreme subwavelength near-field confinement, which allows the local helicity to significantly exceed that of the incident plane wave. Inside the gap, the optical field exhibits opposite helicities at the two metal-dielectric interfaces: the near-field is left-elliptically polarized at the inner core boundary, and right-elliptically polarized at the outer ring boundary. Because the surface normal vectors of these two interfaces are physically opposed, this inverse polarization symmetry drives IFE drift currents that circulate in the exact same angular direction within both the inner metallic core and the outer metallic ring, thereby synergistically amplifying the generated magnetic field.

When examining the higher-order Fabry-Perot modes, the electric field and optical spin density distributions morph into a characteristic two-lobe standing-wave profile along the depth of the gap. Despite this nodal separation, the optical polarization remains completely consistent between the upper and lower lobes. This dictates that the drift currents generated in both the top and bottom sections of the nanoantenna flow in the same direction. Consequently, once these synchronized currents are integrated, they generate a remarkably robust and spatially homogeneous out-of-plane magnetic field ($B_z$) distribution throughout the entire coaxial structure.

Most interestingly, this Fabry-Perot mechanism proves to be highly efficient. When evaluating the first-order FP resonance occurring at $800$ nm for the $200$ nm-thick structure, the generated magnetic field amplitude is directly comparable to the $B_z$ intensity produced by the fundamental LSPR of a thinner coax explicitly tuned to resonate at that exact same $800$ nm wavelength. As shown in **Figure S3**, this striking amplitude parity between two fundamentally distinct resonant modes reveals a profound physical insight: while structural geometry is the primary tool for shifting the resonance to a target wavelength, the maximum achievable magnetic field amplitude at that frequency is ultimately dictated by the intrinsic optical properties of the metal itself. Consequently, once the nanoantenna is geometrically optimized to resonate at a specific frequency, the absolute strength of the IFE magnetization becomes fundamentally a material-driven property. This confirms that higher-order FP modes are not mere secondary artifacts, but exceptionally efficient, geometry-driven channels for localized magnetization.

Collectively, these findings demonstrate that this work not only refines but significantly extends established theoretical frameworks. While a general trend of LSPR-enhanced IFE magnetic fields at longer wavelengths aligns with classical predictions from Hertel's Drude model,[6] our robust numerical modeling, anchored by strict fits of experimental refractive indices, demonstrates that material selection and spatial geometric tuning are paramount for optimizing device performance across distinct spectral ranges. Consequently, this comprehensive optimization library provides a novel, practical pathway toward achieving a highly tunable, multispectral magneto-optical IFE platform.

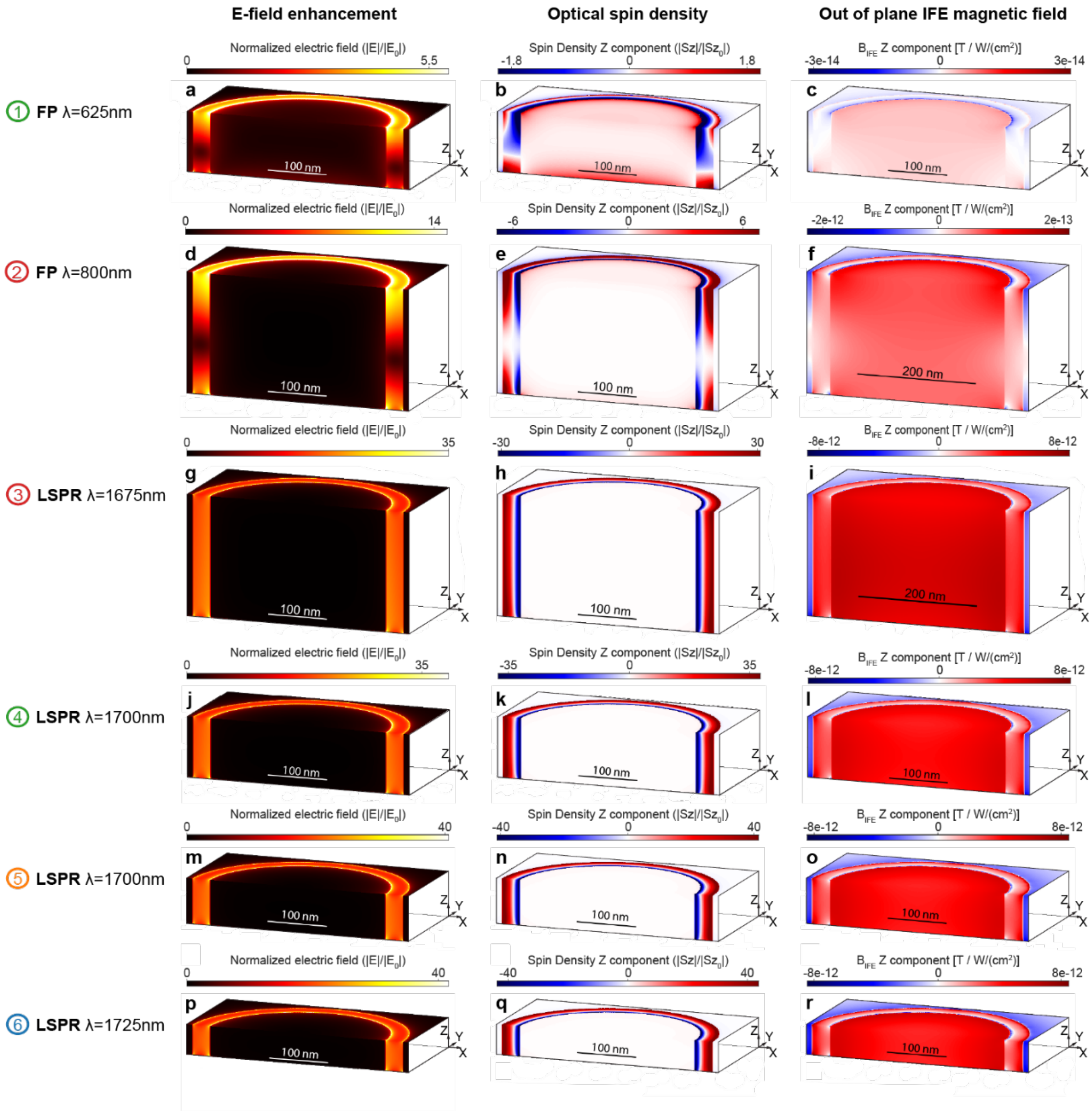

**Figure 4: Near-field spatial mapping of multispectral IFE magnetization.** Cross-sectional maps (in both XY and XZ planes) characterizing the local field distributions within resonant Gold coaxial nanostructures ($D = 300$ nm). For each geometric configuration, three spatial distributions are evaluated: the optical electric field enhancement ($|\mathbf{E}|/|\mathbf{E}_0|$), the local optical spin density ($S_z$), and the resulting out-of-plane IFE magnetic field component ($B_z$). Panels **(a–l)** map these distributions for the fundamental Localized Surface Plasmon Resonance (LSPR) across varying structural thicknesses: $T = 50$ nm (a–c), $75$ nm (d–f), $100$ nm (g–i), and $200$ nm (j–l). Panels **(m–r)** map the distributions corresponding to the first-order Fabry-Perot (FP) resonances, specifically comparing the intermediate thickness $T = 100$ nm (m–o) with the thicker structure $T = 200$ nm (p–r), which exhibits a characteristic two-lobed profile.

## Conclusion

In conclusion, this study establishes a comprehensive and predictive theoretical framework for engineering multispectral all-optical magnetization via the Inverse Faraday Effect. By harnessing the geometric versatility of plasmonic coaxial nanoantenna, we have moved beyond the traditional constraints of gold-centric, narrowband models. Our systematic mapping across a diverse material library, highlighting Aluminum for the deep-UV, and Copper and Silver as highly efficient alternatives in the near-infrared (NIR), demonstrates that the synergistic tuning of material permittivity and structural dimensions enables precise spectral control of both fundamental LSPR and higher-order Fabry-Perot resonances. Furthermore, the striking amplitude parity between distinct modes tuned to the exact same wavelength reveals a profound physical rule: while nanoscale geometry is the primary tool for shifting the resonance to a target frequency, the maximum achievable magnetic field strength is fundamentally governed by the intrinsic optical properties of the selected metal.

Crucially, by resolving the local optical spin density, we have elucidated the underlying near-field mechanics of this phenomenon. The extreme subwavelength confinement within the coaxial aperture generates an inverse polarization symmetry, driving synchronized, non-dissipative drift currents that synergistically amplify the localized magnetic field. As demonstrated, this mechanism becomes exceptionally efficient at longer wavelengths, where the magnitude of the generated out-of-plane static magnetic field ($B_z$) increases by several orders of magnitude.

Ultimately, these findings unlock transformative perspectives for the fields of nanoscale spintronics and fundamental light-matter interactions. The ability to deterministically generate intense, highly localized magnetic fields on subpicosecond timescales entirely through optical means circumvents the spatial and temporal limitations of traditional electromagnets. By coupling these engineered plasmonic nanoantennas to adjacent ferromagnetic or ferrimagnetic thin films, this versatile platform provides a direct, highly efficient pathway toward helicity-dependent all-optical switching (AOS). Merging the ultrafast bandwidth of nanophotonics with the non-volatile robustness of nanomagnetism, this optimized optomagnetic architecture paves the way for the next generation of low-energy, ultra-high-density, and ultrafast data storage devices.

**Acknowledgements**
This work is supported by the ERC grant FemtoMagnet (grant no. 101087709). We acknowledge the financial support from the Agence Nationale de la Recherche (ANR-20-CE09-0031-01, ANR-22-CE09-0027-04 and ANR-23-ERCC-0005), the Institut de Physique du CNRS (Tremplin@INP 2020). M.M. and Y.M. gratefully acknowledge the support of K. C. Wong Education Foundation.

**Availability of data and material**
The materials, data and code underlying the results presented in this paper are not publicly available at this time but may be obtained from the authors upon reasonable request.

**Supporting Information**
Supporting Information is available from the Wiley Online Library or from the author. It contains extended details on the full-wave numerical implementation, mesh convergence, and the effective circularly polarized source. Additionally, it provides the material permittivity models, the analytical derivation of the optical spin density, and supplementary figures illustrating the amplitude parity between LSPR and Fabry-Perot resonances.

**Conflict of Interest**
The authors declare no conflict of interest.

**References**

[1] J. P. van der Ziel, P. S. Pershan, L. D. Malmstrom, *Phys. Rev. Lett.* **1965**, *15*, 190. doi:10.1103/physrevlett.15.190

[2] P. S. Pershan, J. P. van der Ziel, L. D. Malmstrom, *Phys. Rev.* **1966**, *143*, 574. doi:10.1103/physrev.143.574

[3] A. V. Kimel, A. Kirilyuk, T. Rasing, *Laser Photonics Rev.* **2007**, *1*, 275. doi:10.1002/lpor.200710022

[4] A. Kirilyuk, A. V. Kimel, T. Rasing, *Philos. Trans. R. Soc. A* **2011**, *369*, 3631. doi:10.1098/rsta.2011.0168

[5] C. D. Stanciu, F. Hansteen, A. V. Kimel, et al., *Phys. Rev. Lett.* **2007**, *99*, 047601. doi:10.1103/physrevlett.99.047601

[6] R. Hertel, *J. Magn. Magn. Mater.* **2006**, *303*, L1. doi:10.1016/j.jmmm.2005.10.225

[7] R. Hertel, M. Fähnle, *Phys. Rev. B* **2015**, *91*, 020411. doi:10.1103/physrevb.91.020411

[8] R. Sinha-Roy, J. Hurst, G. Manfredi, P. A. Hervieux, *ACS Photonics* **2020**, *7*, 2429. doi:10.1021/acsphotonics.0c00462

[9] X. Yang, Y. Mou, B. Gallas, A. Maitre, L. Coolen, M. Mivelle, *ACS Nano* **2021**, *16*, 386. doi:10.1021/acsnano.1c06922

[10] Y. Mou, X. Yang, M. Vega, et al., *Nano Lett.* **2024**, *24*, 7564. doi:10.1021/acs.nanolett.4c00558

[11] O. H.-C. Cheng, D. H. Son, M. Sheldon, *Nat. Photonics* **2020**, *14*, 365. doi:10.1038/s41566-020-0603-3

[12] B. Talin, V. P. Kaftandjian, L. Klein, *Phys. Rev. A* **1974**, *11*, 648. doi:10.1103/physreva.11.648

[13] J. Deschamps, M. Fitaire, M. Lagoutte, *Phys. Rev. Lett.* **1970**, *25*, 1330. doi:10.1103/physrevlett.25.1330

[14] C. Hareau, X. Yang, M. Sanz-Paz, M. Sheldon, M. Mivelle, *ACS Photonics* **2025**, *12*, 6535. doi:10.1021/acsphotonics.5c01709

[15] X. Yang, Y. Mou, R. Zapata, B. Reynier, B. Gallas, M. Mivelle, *Nanophotonics* **2023**, *12*, 687. doi:10.1515/nanoph-2022-0488

[16] Y. Mou, X. Yang, B. Gallas, M. Mivelle, *Adv. Mater. Technol.* **2023**, *8*, 2300770. doi:10.1002/admt.202300770

[17] A. Nadarajah, M. T. Sheldon, *Opt. Express* **2017**, *25*, 12753. doi:10.1364/oe.25.012753

[18] V. Amendola, R. Pilot, M. Frasconi, O. M. Maragò, M. A. Iatì, *J. Phys. Condens. Matter* **2017**, *29*, 203002. doi:10.1088/1361-648x/aa60f3

[19] M. Melli, A. Polyakov, D. Gargas, et al., *Nano Lett.* **2013**, *13*, 2687. doi:10.1021/nl400844a

[20] R. Zapata, D. Singh, O. Markovic, et al., *ACS Appl. Mater. Interfaces* **2026**. doi:10.1021/acsami.6c07175

[21] A. K. González-Alcalde, X. Shi, V. H. Ortiz, J. Feng, R. B. Wilson, L. T. Vuong, *Nanophotonics* **2024**, *13*, 1993. doi:10.1515/nanoph-2023-0777

[22] O. H.-C. Cheng, B. Zhao, Z. Brawley, D. H. Son, M. T. Sheldon, *Nano Lett.* **2022**, *22*, 5120. doi:10.1021/acs.nanolett.2c00571

[23] V. Karakhanyan, C. Eustache, Y. Lefier, T. Grosjean, *Phys. Rev. B* **2021**, *105*, 045406. doi:10.1103/physrevb.105.045406

[24] S. Parchenko, K. Hofhuis, A. Å. Larsson, et al., *Adv. Photonics Res.* **2024**, *6*, 2400083. doi:10.1002/adpr.202400083

[25] X. Yang, Y. Mou, B. Gallas, S. Bidault, M. Mivelle, *Nanophotonics* **2025**, *14*, 2453. doi:10.1515/nanoph-2025-0096

[26] X. Yang, C. Hareau, M. Mivelle, *Nano Lett.* **2025**, *25*, 11549. doi:10.1021/acs.nanolett.5c02039

[27] V. H. Ortiz, S. B. Mishra, L. Vuong, S. Coh, R. B. Wilson, *Phys. Rev. Mater.* **2023**, *7*, 125202. doi:10.1103/physrevmaterials.7.125202

[28] D. Yoo, D. A. Mohr, F. Vidal-Codina, et al., *Nano Lett.* **2018**, *18*, 1930. doi:10.1021/acs.nanolett.7b05295

[29] D. Yoo, F. Vidal-Codina, C. Ciracì, et al., *Nat. Commun.* **2019**, *10*, 4440. doi:10.1038/s41467-019-12038-3

[30] D. Yoo, N. C. Nguyen, L. Martin-Moreno, et al., *Nano Lett.* **2016**, *16*, 2040. doi:10.1021/acs.nanolett.6b00024

[31] E. Barakat, M. P. Bernal, F. I. Baida, *J. Opt. Soc. Am. B* **2013**, *30*, 638.

[32] F. I. Baida, M. P. Bernal, E. Popov, A. L. Fehrembach, *Opt Express* **2015**, *23*, 21505.

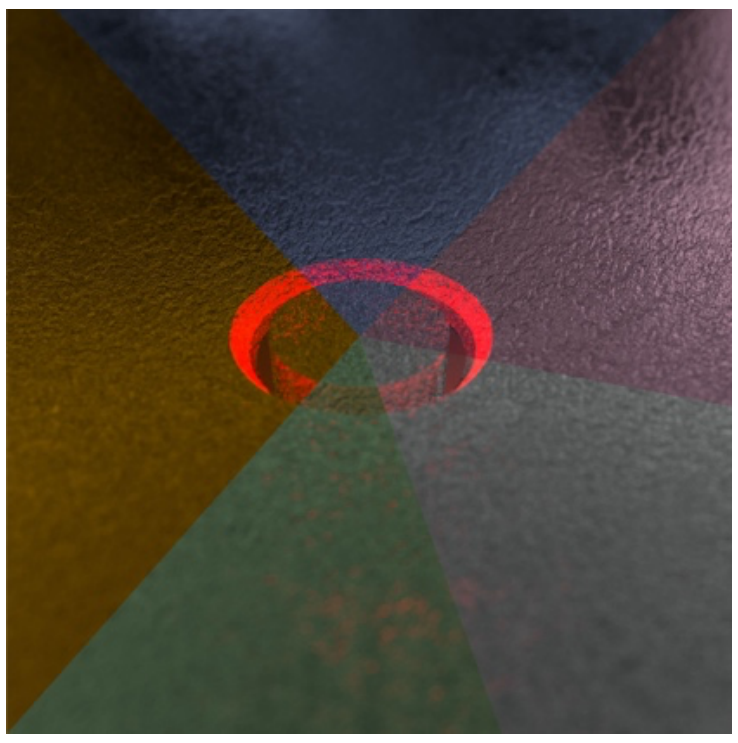

**Table of Contents text:** A comprehensive framework maps the Inverse Faraday Effect across a diverse plasmonic material library using tunable coaxial nanoantennas. Extreme subwavelength confinement of the optical spin density drives synchronized drift photocurrents, enabling precise multispectral optomagnetism. This yields robust deep-UV magnetization in aluminum, and a giant, orders-of-magnitude magnetic enhancement in the near-infrared using gold, silver, or copper.